# Scission of flexible polymers in contraction flow: predicting the effects of multiple passages

Sandeep Garrepally* [1, 2-a)], Stephane Jouenne [1, 2-b)], Peter D. Olmsted[3-c)], Francois Lequeux*[4, 5 -d)].

1) Laboratoire Physico-Chimie des Interfaces Complexes, Bâtiment CHEMSTARTUP, Route Départemental 817, 64170 Lacq, France.
2) TOTAL SA, Pôle d'Etudes et Recherche de Lacq, BP 47, 64170 Lacq, France.
3) Department of Physics and Institute for Soft Matter Synthesis and Metrology, Georgetown University, Washington, DC 20007, United States.
4) Laboratoire Sciences et Ingénierie de la Matière Molle, ESPCI Paris, PSL University, Sorbonne Université, CNRS UMR 7615, F-75005 Paris, France.
5) Laboratoire Physico-Chimie des Interfaces Complexes, ESPCI Paris, 10 rue Vauquelin, F-75231 Paris, France.

**Abstract**

**When injected through a contraction, high molecular weight polymer solutions exhibit a sharp increase of apparent viscosity which originates from stretching polymer chains above a critical extension rate. This chain stretching can also induce polymer scission, which then decreases the extensional viscosity. In practice, the two phenomena are difficult to separate. Moreover, these phenomena are often observed in situation where flow instabilities appear. In order to disentangle the two effects we have measured the pressure-flux relation for polymer solutions passing through a hyperbolic contraction. The ratio of the pressure drop to that of the (Newtonian) solvent has a maximum due to the competition between polymer extension and scission. We find a geometry-dependent relation between the flow rates at which the maximum occurs for successive passages in a given contraction, which appears to be independent of molecular weight, concentration, solvent quality and viscosity, and can be used to predict the scission under successive passes.**

## 1. Introduction

Aqueous polymer solutions in elongational flow have numerous applications in different industries, such as extraction of oil by polymer flooding, polymer melt extrusions, fiber spinning, lubricants and film blowing. High molecular weight polymers melts and solutions can exhibit shear thinning under simple shear flow and hardening under extensional flow.

The phenomenon of shear thinning in polymer solutions has been well studied. It originates from the orientation and disentanglement of chains under flow [1]. Shear flow is a superposition of extension and rotation, in which a polymer chain stretches along the extensional direction, but rotates so that the coil then contracts. This leads to weak stretching and little or no degradation [2, 3]. By contrast, in pure extensional flow the chains stretch and eventually completely uncoil. At macroscopic scales this leads to a large increase of the extensional viscosity [7] de Gennes (1974)] predicted that polymer coils in dilute solution under extensional flow will experience a sudden coil-stretch transition at a critical strain rate $\dot{\varepsilon}_C$ which is inversely proportional to the longest relaxation time of the polymer coil [4].

By increasing concentration the critical strain rate $\dot{\varepsilon}_C$ decreases, as inferred from capillary extensional rheology experiments [5] and from Brownian dynamics simulations with hydrodynamic interactions [6]. Indeed for both dilute solutions and semi dilute solutions, the strain rate for onset of stretching was found to decrease with concentration [7-9]. This was attributed to interactions between polymers during stretching and suggested an ultra-dilute regime where there would be no entanglements or interactions under stretching [5, 10].

The extensional viscosity at high extension rates is primarily due to the resistance offered by the stretching of polymers. The transient extensional viscosity depends on both strain rate and strain. Gupta *et al.* measured the extensional stress growth as a function of strain in dilute and semi dilute polystyrene solutions using a filament stretching device [11]. They found that, while linear shear stress relaxation follows Zimm relaxation due to hydrodynamics, extensional stress growth is better-described by Rouse dynamics. The Zimm-Rouse transformation occurred at a Hencky strain $\varepsilon_H \approx 2$. In this paper we work in the semi dilute un-entangled domain with a geometry whose hencky strain is greater than 2 where the Rouse dynamics are applicable.

As the chain uncoils and extends, hydrodynamic drag between solvent and polymer can lead to scission of polymers for extension rates exceeding a critical rate $\dot{\varepsilon}_f$ when the extensional force approaches or exceeds that required to break a C-C bond (typically nN per chain) [12]. More precisely, scission can be explained from thermal activation assisted by the increase in tension, which decreases the activation energy barrier required for breaking the C-C bond [2].

Scission in extensional flow leads to huge viscosity losses [13], and both stretching and scission depend on flow history and on the precise geometry of extensional flow [14, 15]. In stagnation devices (*e.g.* cross slot or four mill devices) the residence time is considered infinite for the polymers trapped at the stagnation point. The critical strain rate $\dot{\varepsilon}_f$ for scission in this case scales with molecular weight as $\dot{\varepsilon}_f \propto 1/M_w^2$ [2, 16]. This scaling results from concentrating the viscous drag on an extended string of $N$ beads at the center of the chain, where the maximum chain tension $\sigma$ scales as as $\sigma \propto \eta_s \dot{\varepsilon} N^2$ where $\eta_s$ is the solvent viscosity [16].

In contrast, for transient extensional flows there are several observed scalings for the critical strain rate $\dot{\varepsilon}_f \propto M_w^{-b}$, depending on molecular weight, concentration, solvent quality, and whether the flow is laminar or turbulent (Table 1). The weak molecular weight dependence in turbulent flow ($b < 1.3$) is believed to be due to partial stretching. An exponent $b = 2$ would apply for a fully stretched polymer chain, but turbulent flow frustrates this and reduces full extension relative to laminar flow. However, as shown by Table 1, there is not yet a distinct agreement on this topic. Breakage in transient flow has been modelled by several authors [17]. In Ryskin's "yo-yo model " of surface energy accumulation, polymer chains under extension stretch first at the center [18]. The remaining coiled portions at the extremes experience high viscous drag that is balanced by an increased tension at the midpoint. Larson and Magada [19] studied a bead-spring model and found maximum extension at the center of the chain, which is supported by the hairpin model of Nguyen and Kaush [20]. Brett suggested that flow-induced scission leads to a narrower molecular weight distribution, with the decrease in polydispersity due to chain scission from pass to pass suggesting a precise facture pattern [21]. By contrast, Muller suggested that for highly concentrated polymer solutions, scission is more randomized due to entanglements [12]. In dilute solutions tension is exerted by solvent-monomer friction, whereas in semi dilute solutions stress is also transmitted through entanglements. In the latter case the stress per chain which effects degradation is controlled by the entanglement density $c/c^*$ [22], where $c$ is the polymer concentration and $c^*$ is the overlap concentration. To sum-up, there is not yet any clear consensus of the precise mechanism of scission in polymer chains under flow.

Quality of solvent is also an important parameter in chain scission. An increase in solvent viscosity decreases the critical strain rate for the onset of stretching [23]. Odell and Keller proposed a weak dependence for onset of degradation with solvent viscosity, $\dot{\varepsilon}_f \propto \eta_s^{-0.25}$ [24]. Indeed, degradation is more dependent on solvent quality than on solvent viscosity. In good solvents, polymers are more swollen than in bad solvent, and thus more easily stretched by flow gradients. A stretched polymer in poor solvents resemble a pearl necklace (with the 'pearls' comprising small coiled sections of polymers) [25, 26]. The extensional viscosity at a given extensional rate is observed to be lower in bad solvents than in good solvents [9]. Moussa et al. studied the influence of solvent quality on degradation in turbulent flow, in which the motion between two counter rotating eddies involves extensional deformations [27]. They found that the onset for degradation in theta solvents occurred at a lower Reynolds number than in good solvents [28]. Similarly, Odell *et al.* studied polystyrene degradation in cross slot devices in toluene (good solvent) and decalin (theta solvent), finding a lower onset fracture strain rate in the second case. However such comparison is delicate since both relaxation time and c/c* are different in the two systems [13].

Polydispersity influences the critical strain rate for the onset of stretching [8]. Increasing the high molecular weight tail at a given average molecular weight should increase the extensional viscosity [29, 30]. Polydispersity should have a similar influence on degradation (a distribution that enhances the higher molecular weight tail is more likely to undergo scission), although there have been few systematic studies.

For applications such as oil recovery, solutions of flexible polymers are injected into a porous medium, which comprises pores and necks that resemble contraction-expansion geometries. The coil stretch transition of individual polymer chains occurs for a Deborah number $\text{De}_C = 0.5$. However, the spatial distribution of Deborah numbers experienced in porous medium, combined with polydispersity, is complex and will `smear out' the coil-stretch transition [31]. Jouenne et al. observed that scission of polymers in porous medium is controlled by the pressure drop [32]. Several authors have studied scission in porous media by performing experiments on multiple contractions, where they observed that the most significant decrease in the molecular weight distribution occurs during the first contraction and depends on the strain and strain rate of the contraction [33]. So the literature on polymer scission under extensional flow indicates that there exists multiple parameters influencing scission. In this paper, we report a simple way to quantify extensional flow scission that allows to predict scission, in a given geometry, for different solvent qualities and polymer masses.

In this paper, we study polymer scission under extensional flow simply by measuring the pressure drop through a model hyperbolic contraction as a function of mass flow rate. The ratio of the pressure drop of the polymer solution to that of the pure solvent exhibits a maximum for a given flow rate, revealing a crossover between chain stretching and scission. We then study multiple injections to understand the scission. We find a universal relation for predicting the scission after many passages in a given geometry, based on the measurement of the maximum pressure ratio as a function of flow rate through the first contraction, and is valid for whatever the concentration, the molar mass and the solvent quality

## 2. Materials and methods

### 2.1 Polymer solutions

Partially-hydrolyzed polyacrylamide (HPAM) with 30% hydrolysis (Figure 1) was obtained as powder from the Flopaam series (SNF Floerger, Andrezieux, France). The polymers were synthesized by the supplier using radical polymerization and are highly polydisperse, with $M_w/M_n \sim 4 - 5$. Oligomeric polyacrylamide chains (PAM) were also obtained from SNF. The molecular weights obtained are summarized in Table 2.

| Polymer | Solvent | $M_w$ range (MDa) | solvent quality | Concentration regime | b | Flow type | Reference |
|---|---|---|---|---|---|---|---|
| PEO | Water | 2-7 | Good | Dilute | 1 | contraction | [14] |
| Polystyrene | Decalin | 2-4 | Poor | Dilute | 1.3 | Turbulent-contraction | [14, 15] |
| Polystyrene | Dimethylpthalate | 2-1 | Good | Dilute | 0.95 | Turbulent-contraction | [20] |
| HPAM | Water (20g/l NaCl) | 8-12 | Good | Semi-dilute entangled | 1 | Turbulent-contraction | [34] |
| HPAM | Water (20g/l NaCl) | 8-18 | Good | Semi-dilute | 3 | Laminar-contraction | [34] |
| PEO | Water / Water/glycerol (20/80) | 1-8 | Good / Poor | Semi-dilute | 1.22 / 1.42 | Turbulent-contraction | [34] |
| PEO | Water | 1-8 | Good | Semi-dilute | 3.18 | Laminar-contraction | [34] |
| PEO | Water | 0.6-8 | Good | Semi-dilute | 1.7±0.3 | Laminar-contraction | [13] |
| PEO | Water | 2-7 | Good | Dilute | 1.23±0.12 | Turbulent-contraction | [17] |
| PAM | Water | 2-4 | Good | Dilute | 1.27±0.16 | Turbulent | [17] |
| Polystyrene | Decalin | 4-30 | Poor | Dilute | 2 | Steady state Extension-cross slot | [2, 15] |

Table 1: Molecular weight dependence for critical strain rate of onset of degradation found by different researchers and their conditions of degradation. The results summarized above are in fast transient extensional flows where devices consists of an orifice or a contraction. The critical strain rate for onset of degradation scaling with molecular weight as $\dot{\varepsilon}_f \propto M_w^{-b}$. In case of steady state scission, due to the infinite residence time the polymer chains completely unravels and the value of exponent would be $b = 2$ [2, 14].

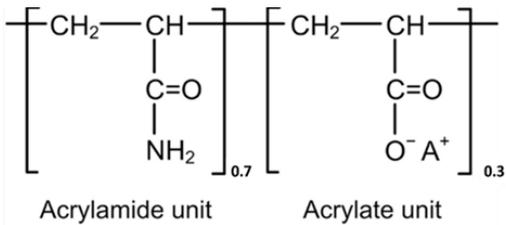

Figure 1: Partially hydrolyzed polyacrylamide (HPAM)

A 0.6% (w/w) salt solution is prepared by mixing 0.0074% of $Na_2SO_4$, 0.0111% of KCl, 0.0745% of $CaCl_2$, 0.0561% of $MgCl_2$, and 0.4724% of NaCl Glycerol and ethanol of analytical grade were obtained from Sigma Aldrich. PAM solution with an oligomeric concentration of 0.9% is prepared in 0.6% salt solution. The PAM solution behaves as a Newtonian solvent in our experimental range of flow rates.

| Polymer grade | $M_w$ (MDa) |
|---|---|
| 3230S | 7 |
| 3330S | 10 |
| 3430S | 11 |
| 3530S | 15 |
| 3630S | 19 |
| PAM | 0.5 |

Table 2: Molecular weights of native polymers of HPAM. IN all cases the polydispersity is $M_w/M_n \sim 4-5$.

Initial mother solutions with HPAM polymer concentration, 0.5 % (w/w) are prepared in a 0.6 % (w/w) salt solution. The mother solution is diluted to desired polymer concentrations using different Newtonian solvents. The viscosity of the solvent is modified by adding glycerol, PAM solution ($M_w = 500$kDa), and ethanol to water. All the polymer solutions prepared in the current work have a fixed salt composition of 0.6 % (w/w). Homogenous mixing is ensured by gently stirring the solutions overnight with a magnetic stirrer. Diluted solutions are then filtered using 5μm nylon screen (we verified that the nylon screen does not degrade the polymer).

The question may arise whether to consider the polyelectrolyte effects *i.e.* the charge interaction between polymers. We follow analysis proposed by Dobrynin *et al.* [35]. The criteria for crossover from polyelectrolyte to non-polyelectrolyte is determined by the concentration of salts $C_s$, and the polymer charge concentration $C_p = c/A$ (where $c$ is the total monomer concentration and $A$ is the number of monomers per effective charge). If $C_s > c/A$, then we do not expect polyelectrolyte effects since solvent ions can screen all of the polymer charges. For our systems the salt concentration $C_s = 215$mM. The polymer concentrations used (~0.08%w/w) results in polymer monomer concentration of order $c$~11mM (0.8/71). For polymers with 30% hydrolysis two charges are separated by 2 monomers ($A = 3$), leading to a polymer charge concentration $C_p = 3.7$mM, which is very low compared to the salt concentration. This confirms that the salt ions are in excess and our solutions are free from polyelectrolyte effects.

## 2.2 Shear characterization and relaxation time

The viscosities $\eta_s$ of Newtonian solvents and $\eta_p$ of the polymer solutions are measured using a ProRheo low shear LS300 rheometer. Viscosities are measured as a function of shear rate in the range 0.1-100 s$^{-1}$. The high sensitivity of this equipment delivers accurate values even at a shear stress of 20µPa. In Figure 2 we show a typical flow curve of polymer solution (viscosity as a function of shear rate). The curve has a Newtonian plateau at low shear rate followed by a shear thinning regime at a critical shear rate which is related to the relaxation time of the polymers. We extract the relaxation time λ of the polymer chain by fitting our experimental data of viscosity using a Carreau model,

$$\eta_p(\dot{\gamma}) = \eta_\infty + \frac{(\eta_0 - \eta_\infty)}{(1+(\lambda\dot{\gamma})^2)^{\frac{(n-1)}{2}}}, \quad (1)$$

where $\eta_\infty$ is the viscosity at infinite shear rate (~$\eta_s$); $\eta_0$ the zero shear viscosity, $\dot{\gamma}$ shear rate and $n < 1$ is the shear thinning index. The rheological data of different polymer solutions are presented in table 4. The relaxation time $\lambda$ of polymer solutions in semi-dilute unentangled polymer solutions obeys Rouse scaling, which follows $\lambda \propto c^{-0.31}M^2$ [36]. In our experiments the relaxation time scales as $\lambda \propto c^{-0.54\pm0.2} M^{1.8\pm0.2}$ (Table 4).

## 2.3 Contraction geometry

Flow through a contraction geometry results in strong extensional flow. A hyperbolic contraction geometry was chosen to obtain a constant rate of stretching $\dot{\varepsilon}_{zz}$ along the length of contraction zone $L_c$ (Figure 3). Certain situations in which the extensional flow does not follow this geometry will be discussed below. Table 3 summarizes the geometry used for the current study. We follow protocol of Nghe *et al.* in fabricating our microfluidic devices [13]:

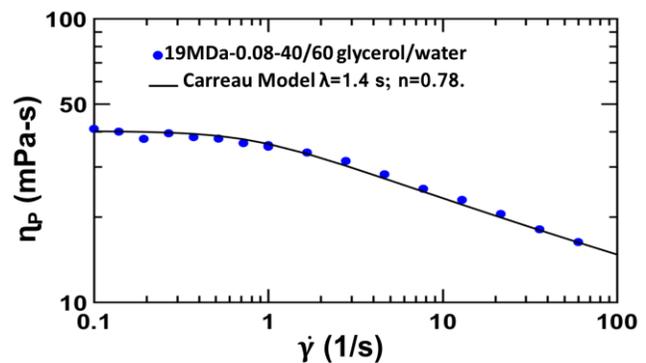

Figure 2: Viscosity of a polymer solution as a function of shear rate obtained using the low shear rheometer. Solid line: Carreau's model fit to obtain the polymer relaxation time.

1. Geometries are fabricated on a silicon wafer in a negative pattern (fabricated by the company "Micro-fluidics")

2. This geometry is transferred from a silicon wafer to a polydimethylsiloxane (PDMS) model.

3. We transferred the geometry from PDMS to microfluidic device using Norland Optical Adhesive 81 ("NOA81") obtained from Thorlabs. NOA81 is a single component liquid adhesive that cures in seconds to a tough, hard polymer when exposed to ultraviolet light.

| Geometry | Entrance width $W_u$ (mm) | Contraction width $W_c$ (mm) | Contraction length $L_c$ (mm) | Channel height $h$ (mm) | Hencky strain $\varepsilon_H$ |
|---|---|---|---|---|---|
| [(5-0.2)/34]$_{0.2}$ | 5 | 0.2 | 34 | 0.2 | 3.21 |

Table 2: Geometry used and its corresponding dimensions (Figure 3).

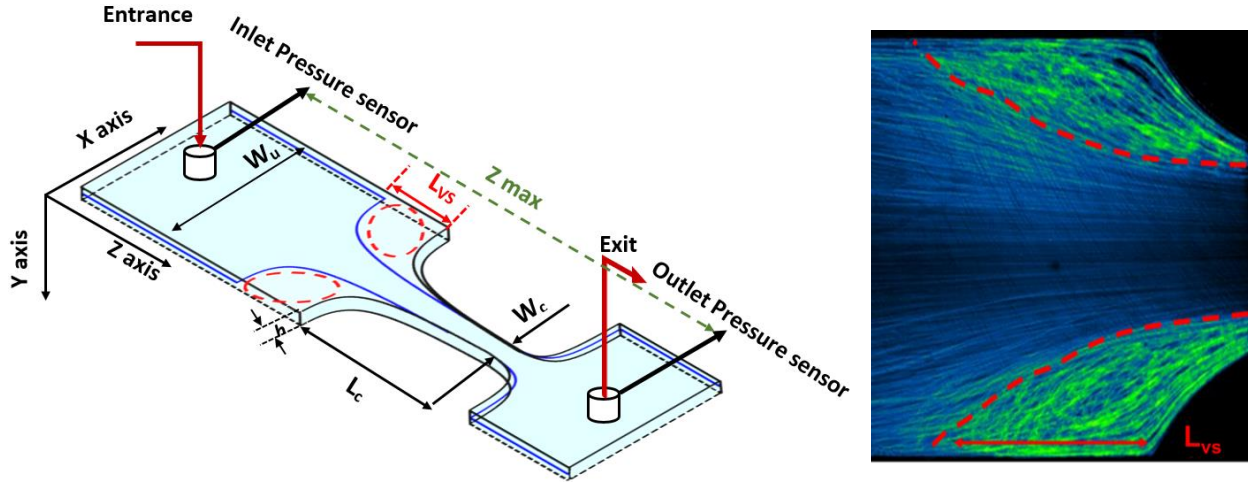

Figure 3: (left) Coordinates and dimensions of the hyperbolic geometry. The red dotted zone shows corner vortices, the black lines denote the boundaries of the cell, and the blue solid line is the boundary of the flow of polymer solution. (Right) Experimentally observed vortices in the contraction geometry, where $L_{vs}$ the length of a vortex is.

## 2.4 Strain rate

We consider fluid moving with velocity $u_z(x, y, z)$ along the Z direction, with width $W(z)$ along X and height h along Y. For a given geometry the extension rate can be estimated as follows.

The imposed uniform volume flux Q in a given geometry, is given by

$$Q = W(z)\, h\, \bar{u}(z) \qquad (2)$$

where

$$\bar{u}(z) = \frac{\iint u_z(x,y,z)\, dx\, dy}{W(z)h} \qquad (3)$$

is the average velocity over height and width at a given $z$. The average extensional rate or the equivalent rate of stretching is given by

$$\bar{\dot{\varepsilon}}(z) = \frac{d\bar{u}(z)}{dz} = \frac{Q}{h}\frac{d}{dz}\left(\frac{1}{W(z)}\right) \qquad (4)$$

For a hyperbolic profile, the width decreases as one passes along the contraction with $(z) = \frac{A(z)}{z}$, where A(z) is the cross-sectional area at a given z. For a contraction with a width Wc located at Z=0 and the start of the contraction with a width Wu located at z=Lc, we find

$$\bar{\dot{\varepsilon}} = \frac{Q}{L_c h}\left(\frac{1}{W_c} - \frac{1}{W_u}\right) \qquad (5)$$

The total accumulated (Hencky) strain undergone by the polymer at the throat of contraction is given by

$$\varepsilon_H = \ln\left(\frac{W_u}{W_c}\right). \qquad (6)$$

The dimensions for our geometry are given in Table 2.

| Name | Molecular weight $M_w$ (MDa) | Polymer concentration %(w/w) | Solvent Composition %(w/w) | n | $\lambda$ (s) | c* %(w/w) | c/c* | $\eta_s$ (mPa-s) | $\eta_p$ (mPa-s) | $Wi_c$ |
|---|---|---|---|---|---|---|---|---|---|---|
| 7MDa-0.08/40SG | 7 | 0.08 | 40% Glycerol | 0.88 | 0.21 | 0.043 | 1.86 | 3.18 | 14 | - |
| 10MDa-0.08/40SG | 10 | 0.08 | 40% Glycerol | 0.86 | 0.31 | 0.035 | 2.28 | 3.18 | 17.75 | - |
| 11MDa-0.08/40SG | 11 | 0.08 | 40% Glycerol | 0.85 | 0.4 | 0.032 | 2.5 | 3.18 | 21.35 | 15±3 |
| 15MDa-0.08/40SG | 15 | 0.08 | 40% Glycerol | 0.8 | 0.86 | 0.029 | 2.76 | 3.18 | 32.5 | 22±5 |
| 19MDa-0.08/40SG | 19 | 0.08 | 40% Glycerol | 0.76 | 1.41 | 0.026 | 2.31 | 3.18 | 42.8 | 15±1 |

| | | | | | | | | | |
|---|---|---|---|---|---|---|---|---|---|
| 19MDa-0.06/40SG | 19 | 0.06 | 40% Glycerol | 0.86 | 0.96 | 0.026 | 3.85 | 3.18 | 20.37 | 17±2.5 |
| 19MDa-0.1/40SG | 19 | 0.1 | 40% Glycerol | - | - | 0.026 | 3.08 | 3.18 | - | - |
| 19MDa-0.08/0SG | 19 | 0.08 | Water | 0.74 | 0.63 | 0.018 | 4.44 | 0.9 | 18.6 | 19±3 |
| 15MDa-0.08/0SG | 15 | 0.08 | Water | 0.8 | 0.43 | 0.022 | 3.63 | 0.9 | 13.1 | 21±4 |
| 11MDa-0.08/0SG | 11 | 0.08 | Water | 0.81 | 0.178 | 0.03 | 2.67 | 0.9 | 7.4 | 15±3 |
| 19MDa-0.08/0,9SP | 19 | 0.08 | 0,9% PAM | - | - | - | - | 3.21 | - | - |
| 19MDa-0.08/15SE | 19 | 0.08 | 15% Ethanol | 0.84 | 0.36 | 0.045 | 1.78 | 1.44 | 8.8 | 14±3 |
| 15MDa-0.08/15SE | 15 | 0.08 | 15% Ethanol | 0.89 | 0.22 | 0.5 | 1.6 | 1.44 | 6.7 | 13±4 |
| 11MDa-0.08/15SE | 11 | 0.08 | 15% Ethanol | 0.92 | 0.15 | 0.064 | 1.25 | 1.44 | 4.7 | 11±3.5 |

Table 4: Nomenclature to specify the type of polymer solutions for a given experiment. Here $n$ is the shear thinning index, $\lambda$ the relaxation time from the Carreau model. c* is the critical (overlap) concentration that separates the dilute and semi-dilute regimes and is calculated as the inverse of intrinsic viscosity. the geometry used in the experiments is **[(5-0.2)/34]**$_{0.2}$. The salinity of all solutions is 0.6% (w/w) salt composition as described in section 2.1.

### 2.5. Rheological data of polymers:

We use the following nomenclature to identify the molecular weight, concentrations of polymer and solvent, and geometry: **"7.2 MDa-0.08/40SG6"** signifies **"$M_w$- Polymer Concentration %(w/w)/ Solvent Composition (w/w)"**. Polymer solutions are summarized in Table 4.

### 2.6 Optical visualization

Polymer solutions in contractions are well known to exhibit vortices above a critical flow rate [37-40]. A typical situation is shown in Figure 3, where the red dotted ovals represent the zone of vortices. Polymer solutions for fluorescence imaging are prepared by mixing polystyrene spherical beads (Thermofischer) of 10μm diameter at a concentration of 3.6*10$^6$ beads/ml. The beads are loaded with a fluorescent dye with spectra in the visible range (absorption at a wavelength of 490 nm and emission at 520nm). Fluorescence imaging is carried out using a Nikon microscope with broadband illumination. Filters inbuilt into the microscope select the range of illumination. Flow visualization is performed using a 2X objective. A high speed camera is placed in-line to record the flow at a rate of 50 frames per sec with a high exposure time of 60-70s.

### 2.7 Single injection experiments

In the single injection experiments (Figure 4) pressure sensors (ELVESYS, Paris, France) are placed at the entrance and exit. The pressure sensors have ranges of either 0-7*10$^5$Pa (±10$^3$Pa) or 0-2*10$^5$Pa (±0.3 *10$^2$Pa). The choice of range limits the sensitivity and we choose the pressure sensor for each experiment.

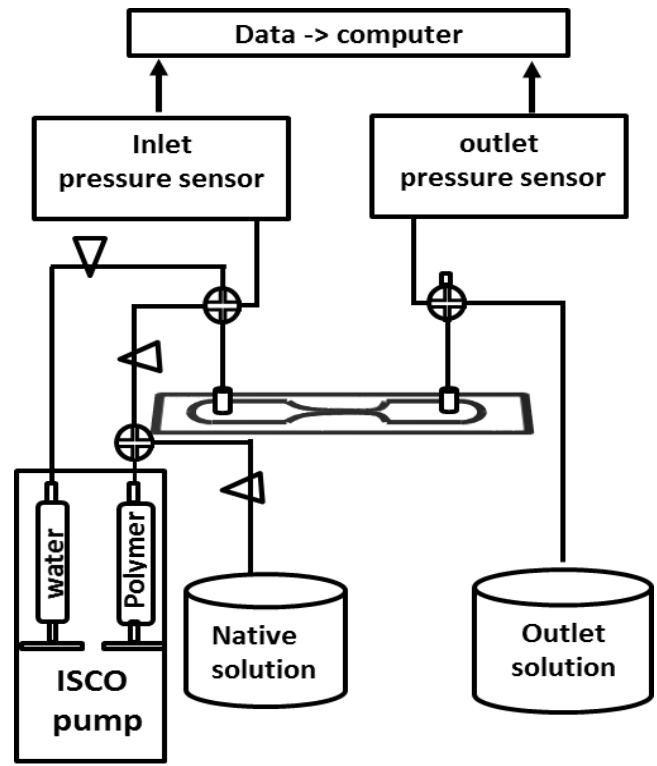

Figure 4: Single injection experimental device including the injection system, the tubing, the micro-model, the collecting bottle and pressure sensors at entrance and exit.

First the solvent is injected through the contraction at different flow rates $Q$ to measure the pressure difference $\Delta P_s(Q)$. The pressure drop due to solvent increases linearly with flow rate (Figure 5), $\Delta P_s(Q) = R_s Q$, where the resistance factor $R_s$ depends on the solvent viscosity and the geometry. The Reynolds number is estimated as $\text{Re} = \frac{\rho \bar{u}(z) d_h}{\eta_s}$, where $d_h = \frac{4A(z)}{2(h+W(z))}$, $\bar{u}(z)$ is the average velocity calculated as in Equation 3, and ρ is the solvent density.

In this paper we work with solvents at Reynolds numbers<150. Deviation from linearity of pressure with flux are observed for Reynolds number (>80), this deviation results from the flow field deformation due to inertia at the exit of contraction. Therefore we neglect these deviations and extrapolate the linear fit of pressure due to solvent. However in case of polymer solutions the Reynold's number is always small (<80) and therefore we do not observe these deviations.

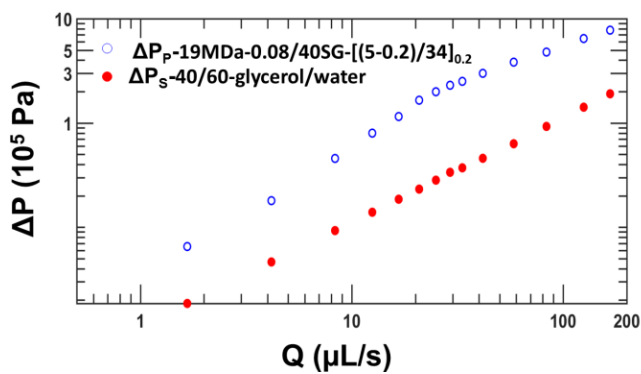

Figure 5: Typical pressure drop $\Delta P_p(Q)$ for a polymer solution compared with that for the solvent, $\Delta P_s(Q)$.

Next we measure the pressure difference $\Delta P_p(Q)$ for the polymer solutions in Table 4 and calculate the ratio of the pressure drops between polymer solution and pure solvent at the same flow rate,

$$P_R(Q) = \frac{\Delta P_P(Q)}{\Delta P_s(Q)}. \qquad (7)$$

A typical plot of pressure drop ratio $P_R(Q)$ has a maximum, as shown in Figure 6, at values of pressure and flux respectively $P_R^{max}$ and $Q^{max}$.

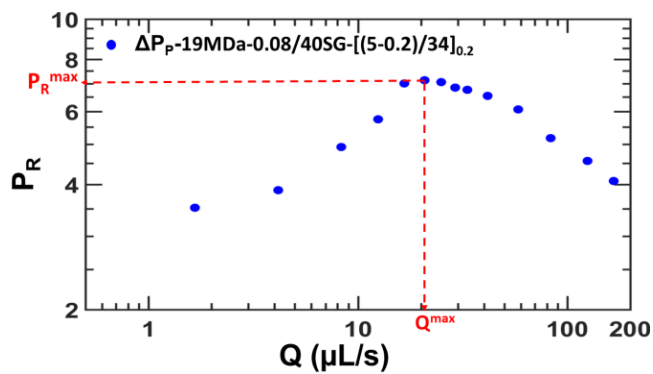

Figure 6: Pressure drop ratio curve of a given polymer solution. The maximum occurs of at pressure $P_R^{max}$ and flow rate $Q^{max}$.

## 2.8 Multiple passage experiments

Figure 7 illustrates a double passage experiment. In step 1 we inject the native solution at different flow rates $Q_1$ and measure the pressure drop ratio curve as explained in the previous section. The blue solid line in the first graph of the Figure 7 is the first injection pressure drop ratio curve having a maximum at $Q_1^{max}$. At the outlet the solution is collected at a flow rate $Q_1$. In step 2 the collected $Q_1$ outlet solution is injected for a second time at different flow rates $Q_2$ to measure the second pressure drop ratio curve, as shown by the orange dotted line with a maximum at $Q_2^{max}$. We observe that this maximum depends on the Initial flow rate of collection $Q_1$ and $Q_1^{max}$. A given multiple injection curve is denoted as $P_R (Q_N; Q_{N-1})$ where $Q_N$ is the flow rate at the $N^{th}$ passage and $Q_{N-1}$ is the flow rate at which the polymer was degraded during the N-1$^{st}$ passage.

Multiple injection experiments were performed for successive injections and collections,. In this experiment a pressure drop curve $P_R(Q_N; Q_{N-1})$ depends on the entire injection history ($Q_1$, $Q_2$, … $Q_{N-1}$). The experimental procedure is similar to the double injection experiments (Figure 8)

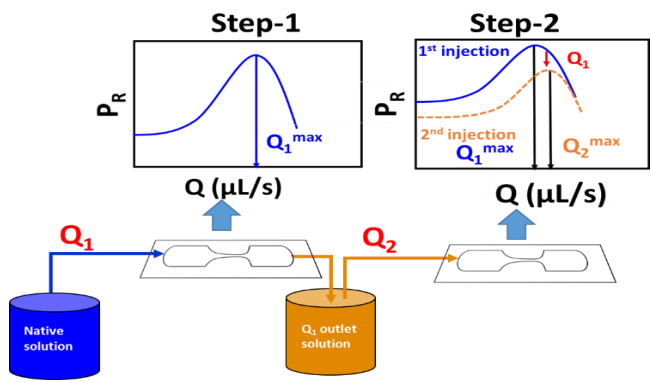

*Figure 7: Double injection technique (Steps 1 and 2).*

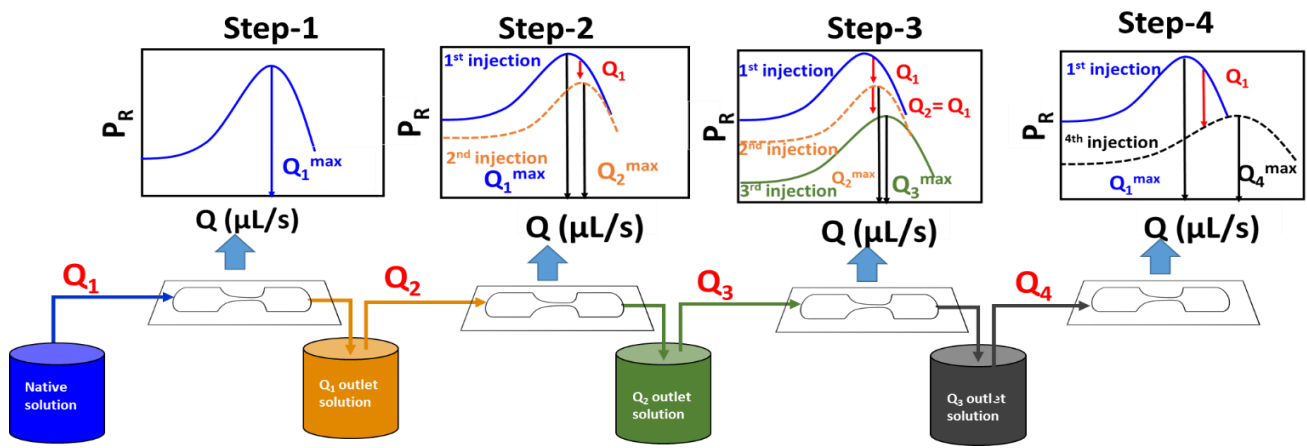

Figure 8: Multiple injection experiments. The blue solid curve is the first injection pressure drop ratio curve $P_R(Q_1)$ with maximum at $Q_1^{max}$. The orange dotted curve is the second injection $P_R(Q_2; Q_1)$ with maximum at $Q_2^{max}$; the green solid curve is the third injection pressure drop $P_R(Q_3; Q_2)$ with maximum at $Q_3^{max}$, and the black dotted line is the fourth injection pressure drop ratio curve $P_R(Q_4; Q_3)$ with maximum at $Q_4^{max}$.

## 3. Results and discussion
### 3.1. Solvent-quality and Concentration domain

As explained in the introduction, the quality of the solvent and the concentration regime both affect chain scission. The relative increase of viscosity due to added polymer is expressed by the *specific viscosity* $\eta_{sp}$,

$$\eta_{sp} = \frac{\eta_p - \eta_s}{\eta_s}. \tag{8}$$

The *reduced viscosity* is the ratio of specific viscosity and polymer concentration $c$,

$$\eta_r = \frac{\eta_{sp}}{c}. \tag{9}$$

With no polyelectrolyte effects the specific viscosity depends on concentration according to the Huggins equation,

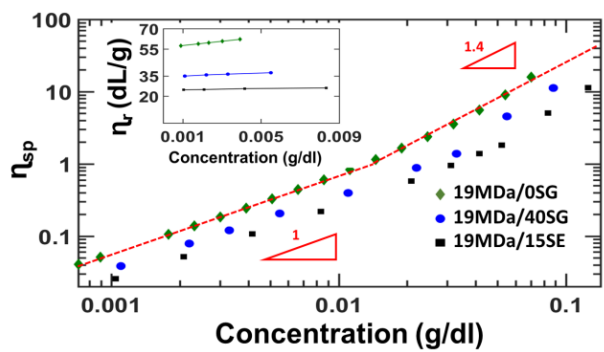

Figure 9: Specific viscosity as a function of concentration for molecular weight 19MDa in three different solvents. Scaling 1 denotes the dilute regime, while scaling 1.3 denotes the semi-dilute unentangled regime. Inset: reduced viscosity as a function of concentration.

$$\eta_r = [\eta] + k_H [\eta]^2 c^2, \tag{10}$$

which defines the *intrinsic viscosity* $[\eta]$ and the Huggins parameter $k_H$, which originates from polymer-polymer interactions. Figure 9 shows the specific and reduced viscosities as a function of concentration for molecular weight 19MDa (Table 5).

| Name | Solvent | $[\eta]$ (dl/g) |
|---|---|---|
| 0SG | Water | 0.0056 |
| 40SG | 40/60 glycerol/water | 0.0034 |
| 15SE | 15/85 ethanol/water | 0.00225 |

Table 5: Intrinsic viscosity and for 19MDa polymer in three different solvents.

The intrinsic viscosities (Table 5) depends on the solvent and obey $[\eta]_{0SG} > [\eta]_{40SG} > [\eta]_{15SE}$. The Fox-Flory equation predicts that the intrinsic viscosity is proportional to the hydrodynamic volume occupied by the polymer $[\eta] \propto V_h \propto R_g^3$ [35]. In good solvents, monomer-monomer repulsion swells the polymer coils, while in poorer solvents a decrease of coil size can be achieved due to a decrease in monomer-monomer repulsion (glycerol) or by disfavoring the monomer-solvent interaction (ethanol). From the intrinsic viscosity values we conclude that the three solvent are good solvent for HPAM. The case of water is known [41]. Glycerol reduces the solvent quality – as compared to water - in a manner similar to a salt, which screens polymer charges and leads to a decrease in the polymer coil size [22, 42, 43]. Ethanol also reduces even more the solvent quality [23].

Figure 9 shows two expected concentration scalings of the specific viscosity with concentration. $\eta_{sp} \propto cM_w^{0.7}$ in the dilute regime where relaxation dynamics best follow the Zimm model [44], while the Rouse model ($\eta_{sp} \propto c^{1.3}M_w^2$ for good solvents, $\eta_{sp} \propto c^2 M_w^2$ for poor solvents) may better describe the semi-dilute unentangled regime [36], which we study here. We observe a concentration scaling $\eta_{sp} \propto c^{1.4\pm0.2}$ consistent with the expected good solvent semi-dilute unentangled regime.

## 3.2. Influence of Molecular weight and Concentration

The aim of this study is to quantify degradation during polymer elongation. The mechanical resistance to polymer stretching is reflected in the increase of the pressure drop ratio (Figure 6). As the flow rate (and thus extension rate) increases the polymer chains stretch, which increases the pressure drop ratio. Stretched polymer coils are susceptible for scission, which then decreases the pressure drop ratio. The competition between elongation and scission leads to the maximum pressure drop ratio $P_R^{max}$ at a flow rate $Q^{max}$.

Figure 10(a) shows the pressure drop ratio for different molecular weights. For smaller molecular weights the pressure drop ratio decreases and the maximum pressure drop ratios occurs at higher flow rates as expected.

We normalize each pressure drop curve both in pressure and in flux by respectively its corresponding maxima $P_R^{max}$ and $Q^{max}$ (Figure 10(b)). The superposition is effective for $Q > Q^{max}$ but fails for $Q < Q^{max}$. We attribute the failing at low flux to the dependence of extensional and shear viscosities on molecular weight are not the same, and to the fact low flow rate is dominated by simple shear. The good superimposition at large shear rate indicates that degradation has the same dependence on flow rate as does the extensional viscosity, which make sense because each of them is related to intra-chain tension.

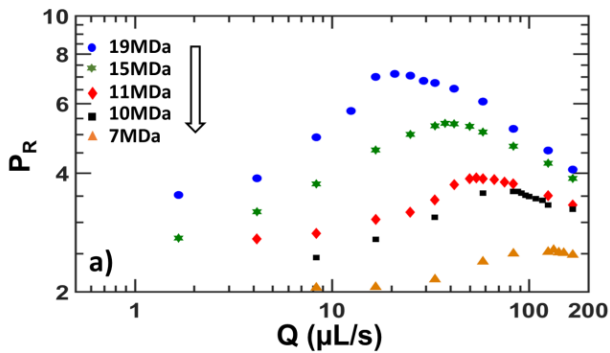

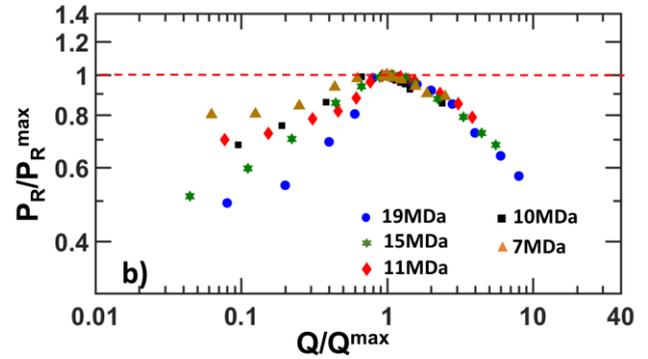

Figure 10 : (a) Pressure drop ratio curve for varying molecular weight (Mw) in 40/60 glycerol water mixture and (b) pressure drop ratios and flow rates normalized by their respective maxima.

The molecular weight and concentration dependence for $P_R^{max}$ and $Q^{max}$ in 40/60 glycerol-water (Figure 11) follow :

$$P_{Rmax} \propto M_w^{1\pm0.2} c^{0.7\pm0.2}$$
$$Q^{max} \propto M_w^{-1.8\pm0.2} c^{-0.7\pm0.2}$$
(11, 12)

A similar scaling was found by Nghe *et al.*, but for the onset of degradation. The authors quantified degradation of semi-dilute PEO solutions in a contraction by the loss of viscosity [13]. Their study suggested a critical strain rate $\dot{\varepsilon}_f$ for onset of degradation, which scales as $\dot{\varepsilon}_f \propto M_w^{-1.7\pm0.3} c^{-0.7\pm0.3}$.

## 3.3. Viscoelastic instabilities

Viscoelastic instabilities easily occur for polymer solutions flowing through contraction devices, often via vortices forming at the entrance. Rothstein and McKinley [45, 46] showed that the onset of vortices occurs at a critical Deborah number, defined as the ratio of the polymer relaxation time λ to the residence time τ:

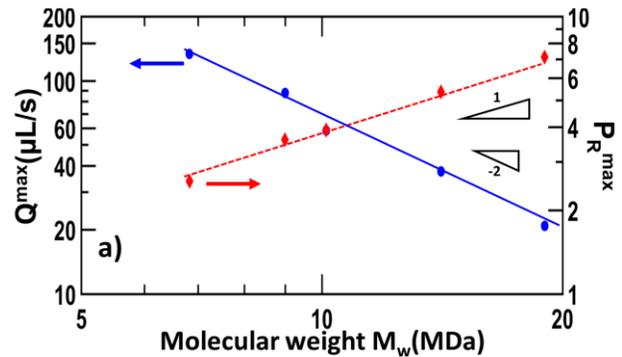

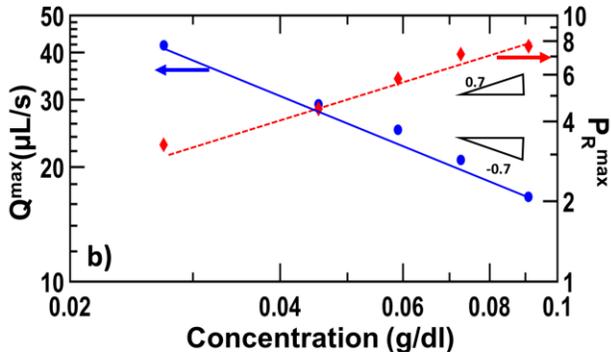

Figure 11: (a) Molecular weight and (b) concentration dependences of $P_R^{max}$ and $Q^{max}$ in 40/60 glycerol-water. The straight lines have the slopes indicated.

$$De = \frac{\lambda}{\tau} \quad (13)$$

They observed three different domains of vortices: lip vortices, corner vortices and unstable vortices. The hyperbolic geometry eliminates lip vortices but allows corner vortices with a well-defined length and circulation, similar to those observed in Rothstein and McKinley [45, 46]. Cloitre *et al.*, suggested that these vortices result from the balance between extensional and shear forces [47]. Hinch suggested that vortices could influence the complete stretching of polymer and/or possibly hold the stretch of polymers for longer times than that possible in laminar flows [48]. We find that the vortex length $L_{vs}$ increases with flow rate; when it reaches the device dimensions, vortices become unstable or chaotic. We suggest that this is likely due to the resulting confinement.

HPAM solutions in water-glycerol mixtures form vortices with length $L_{vs}$ increasing as a function of flux $Q$. On the other hand, HPAM solutions in the ethanol-water solvent with poorer solvent quality (Sec. 3.1) do not form vortices prior to chain scission. We speculate that the poorer solvent quality of ethanol-water leads to a smaller ratio of extensional thickening relative to the shear viscosity, which suppresses vortex formation.

The occurrence of vortices increases the effective contraction length and promotes early stretching of the polymer chains (Figure 3). Hence the strain rate experienced by the solution is lower than that specified by the hyperbolic geometry. To estimate the true strain rate, we increase the contraction length $L_c$ in equation 5 by the vortex length $L_{vs}$ to obtain a modified "true" extension rate

$$\dot{\varepsilon}_T = \frac{Q}{(L_c + L_{vs})h}\left(\frac{1}{W_c} - \frac{1}{W_u}\right). \quad (14)$$

In Section 3.4 below we use the true extensional rate $\dot{\varepsilon}_T$.

## 3.4. Flow conditions at the pressure drop ratio maximum

The extensional strain rate in a hyperbolic geometry is calculated using equation 14. Although the pressure maximum is not correlated with vortex formation, a vortex will reduce the extension rate due to earlier stretching of the chains at the (upstream) beginning of the vortex.

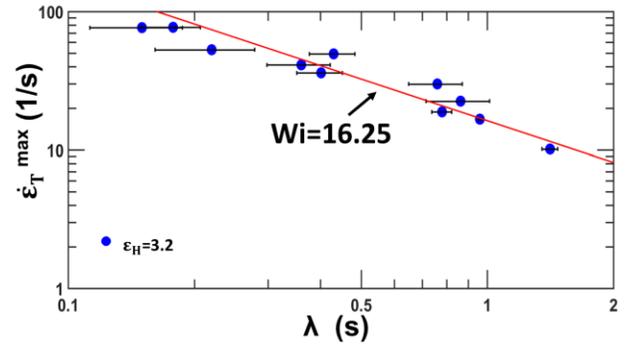

Figure 12: Comparison of maximum extension rate as a function of polymer relaxation time for all samples (with all three solvents). The data are fit by a critical Weissenberg number $Wi_c = 16.25$.

The pressure maximum is associated with chain scission, and the attainment of either (a) a critical tension necessary to rupture a C-C bond (of order pN/chain), or (b) a chain tension sufficiently close to the rupture tension for long enough time to allow thermal activation of rupture [2]. A critical tension will be set by a critical stress and its corresponding extension rate, while activation will depend on the residence time at high chain tension. Hence the pressure maximum could be determined by either a critical Weissenberg number (to exceed the rupture tension)

$$Wi = \dot{\varepsilon}\lambda \quad (15)$$

or a critical Deborah number (to allow thermal activation during extension)

$$De = \frac{\lambda}{\tau} = \frac{\dot{\varepsilon}}{\varepsilon_H}\lambda. \quad (16)$$

However different geometries and extension rates are required to discriminate between these two criteria, which is not in the scope of the current paper.

The pressure maximum is not a direct indicator of the tension during chain scission, since the pressure drop ratio combines shear flow and extensional flow as explained by Cloitre *et al.*[47].

In Figure 12 we show the true extension rate (proportional to Wi) as a function of the relaxation time. Since in the current paper we use only one geometry the Weissenberg number as in Eq.15 and the Deborah number in Eq. 16 would give us similar understandings, In order to differentiate weather chain rupture occurs instantaneously during transient stretching which is controlled by the Weisenberg's number, rather than gradually as in an activated process controlled by the Deborah number one needs further controlled experiments at different Hencky strains.

Vortex lengths were measured on only selected polymer solutions, so for further analysis we use flux as a measure of extension rate.

## 3.5. Double passage

We next show that double injection allows us to follow polymer scission in detail. We first inject a solution with a specific flux $Q_1$ and collect the output of the solution. We then inject this collected output a second time at flux $Q_2$ to measure the pressure drop ratio curve $P_R$ ($Q_2$; $Q_1$) (Figure 13).

The maximum of the second passage pressure drop ratio curve $P_R$ ($Q_2^{max}$; $Q_1$) shifts to higher flux and lower pressure drop ratio in comparison to the first passage pressure drop ratio curve $P_R$ ($Q_1$). When the collection flow rate $Q_1$ is increased, the maximum pressure ratio $P_R^{max}$ of the second passage shifts to lower values and $Q_2^{max}$ shifts to higher values. This is consistent with the finding (Eqs. 11 and 12), that for decreasing molecular weight the maximum of the pressure drop ratio curve shifts to lower values of $P_R^{max}$ and higher $Q^{max}$. This confirms the presence of scission, so that the maximum of the re-injected curve provides information about polymer degradation, with the help of scalings in Eqs. 11 and 12.

For the native solution 19 MDa-0.08/40SG-[(5-0.2)/34]$_{0.2}$ we find $Q_1^{max}$ =20.8 µL/s (Fig. 13). For solutions pre-degraded (collected) at flow rates $Q_1$>>$Q_1^{max}$ (*i.e.* repectively $Q_1$=83.3 µL/s; 125 µL/s; 167 µL/s) the maximum of the pressure occurs at a flow rate nearly equal to the degradation flux, *i.e.* $Q_2^{max} \cong Q_1$. This suggests that all polymers capable of degrading at extension rate corresponding to $Q_1$ have been degraded as illustrated in Figure 14. For polymer solutions pre-degraded at flow rates $Q_1$<<$Q_1^{max}$ (*i.e.* $Q_1$=8.3 µL/s; 4.2 µL/s ;), we find $Q_2^{max} \cong Q_1^{max}$, suggesting weak degradation.

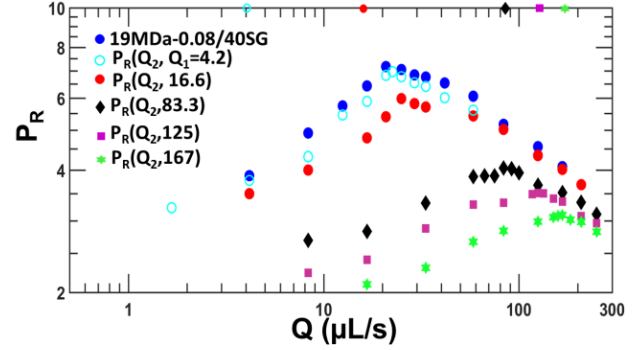

Figure 13: Pressure drop ratio curves of polymer solutions for a second pass at flux $Q_2$ after pre-degradation from a first injection at flux $Q_1$. The first injection is shown as symbols along the upper abscissa.

Similar experiments were done using different molecular weights and solvents (Table 4). The relative position of $Q_2^{max}$ versus $Q_1$ and $Q_1^{max}$ suggests the master curve shown in Figure 15. Here we plot the maximum of the second pressure drop ratio curve $Q_2$ normalized by the maximum of its respective non-degraded solution (*i.e.* $Q_2^{max}/Q_1^{max}$), as a function the normalized pre-degrading flow rate $Q_1/Q_1^{max}$. In this representation, the two asymptotic regimes described above can been observed. For flow rate of the first injection below the maximum pressure flow rate, *i.e.* for $Q_1/Q_1^{max}$ <<1, we see that $Q_2^{max}/Q_1^{max}$ tends toward 1, indicating an absence of polymer scission. For flow rate above the maximum pressure flow rate, *i.e.* $Q_1/Q_1^{max}$ >> 1, we find $Q_2^{max}/Q_1^{max} \approx Q_1/Q_1^{max}$ indicating a saturation scission regime, *i.e.* $Q_2^{max} = Q_1$.

Moreover, for the present contraction all the data collapse onto a master curve, for any molecular weight, concentration above but close to c* and solvent quality. This is the major result of our study, which indicates a general relation for the relative position of the flow rate at the maximum pressure rate between the first and the second passage, for a given contraction geometry, whatever is the solvent quality, concentration and the molecular weight.

This relation can be described by

$$\frac{Q_2^{max}}{Q_1^{max}} = f\left(\frac{Q_1}{Q_1^{max}}\right) = \left[1 + \left(\frac{Q_1}{Q_1^{max}}\right)^a\right]^{\frac{1}{a}}. \qquad (17)$$

Here exponent *a* corresponds to the inverse of the sharpness of the crossover regime.

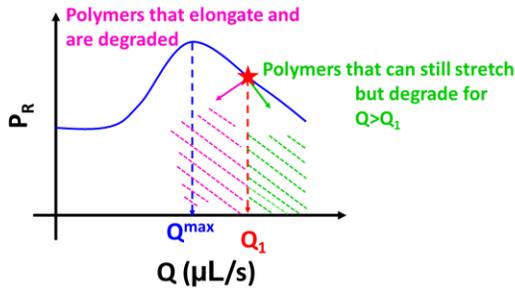

Figure 14: Illustration of the scission at flow rates above $Q^{max}$. Consider a first passage for $Q_1 > Q_1^{max}$ (Fig.14). In the second passage a maximum occurs for flow rates $Q_2^{max} = Q_1$, as the polymers that stretch for flow rates $Q < Q_1$ (pink shaded portion of Fig.14) have already been degraded and do not contribute for extension while polymers non-degraded and contribute for extension at $Q_1$ would only degrade for $Q_2 > Q_1$.

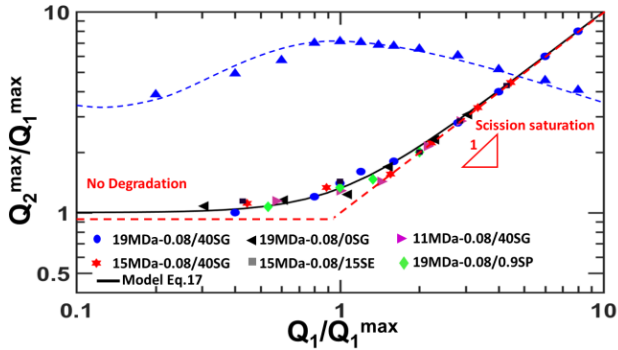

Figure 15: Universal curve for the evolution of pressure drop ratio curve from successive injections for the model geometry **[(5-0.2)/34]**$_{0.2}$. Here, $Q_1$ is the first injection flux, $Q_1^{max}$ is the flux at the pressure maximum for un-degraded material, and $Q_2^{max}$ is the flux at the pressure maximum for material re-injected after being degraded at flux $Q_1$ in the first injection. The solid line shows a fit to equation 17 with exponent a=2.5. Blue triangles are the pressure drop ratio curve as a function of flow rate normalized of a particular polymer and the blue dashed line is a visual for eye.

From the previous scalings between flux maximum and molecular weight (Eq. 12) we can derive the relation between the molecular weights of non-degraded and following passage. Indeed, the molar mass *M* of the solution is characterized by $Q^{max}$. For a non-degraded polymer solution the relation between maximum flux and molecular weight is given by

$$Q_1^{max} = k M_{inlet}^{-2}, \quad (18)$$

where the constant *k* is specific for a given polymer solution and contraction geometry.

When the solution is injected for the first time at given flow rate and the collected outlet solution molecular weight is characterized using the second injection. The relation between maximum flux of second passage and molecular weight of outlet solution is given by

$$Q_2^{max} = k M_{inlet}^{-2} \quad (19)$$

Upon substituting Eqs.18, 19 in evolution of the flux maxima Eq.17 and after due simplification we get the relation between the inlet and the outlet molar mass :

$$M_{out} = M_{in}\left[1 + \left(\frac{M_{in}^2 Q}{k}\right)^a\right]^{-\frac{1}{2a}} \quad (20)$$

The outlet molecular weight is considered to be apparent since in the given molecular weight distribution, polymers contributing for extension and degradation only corresponds to the higher average molecular weights like $M_z$, $M_{z+1}$ ... Eq.12 is the relation of flux maximum with weight average molecular weight ($M_w$), therefore the molecular weight of outlet solution thus calculated in Eq.20 is apparent and cannot be considered as an absolute value.

We plot the apparent outlet molecular weight of Eq.20 with respect to the injected molecular weight (Figure 16). In the limit of a vanishing flux, where one expects no degradation the outlet molecular weight is same as that of the injected molecular weight (linear red line). For a given flow rate the degradation starts to deviate from this linear behavior at a critical molecular weight specific for that flow rate, and then degradation saturates, suggesting that any polymer having molecular weight above this critical value will undergo degradation.

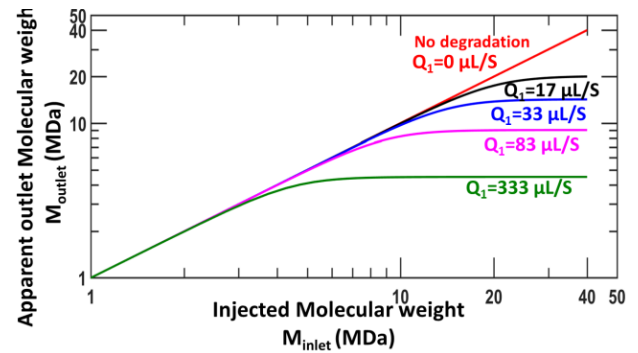

Figure 16: Apparent outlet molecular weight as a function of injected molecular weight at different flow rates, as inferred from equations 12 and 17.

### 3.6. Multiple passages

We next generalize this approach to many passages. Figure 17 summarizes the results for a fluid subjected to four successive injections. In each case we choose $Q_1 = Q_2 = Q_3 = 33.3 \mu L/s > Q_1^{max}$. After each injection the maximum shifts to lower pressure values and higher flow rates, which suggest additional degradation from pass to pass. However, the amount of degradation is not the same from pass to pass.

The maximum of the native solution 19 MDa-0.08/40SG-[(5-0.2)/34]$_{0.2}$ is at $Q_1^{max}$ =20.8 µL/s. The flow rate chosen for the first collection is $Q_1$=33.3 µL /s > $Q_1^{max}$, which thus induces significant degradation. The flow rates chosen for subsequent collections were the same as in the first passage, which in each case is smaller than the maximum for that passage (i.e. $Q_2 < Q_2^{max}, Q_3 \ll Q_3^{max}$).

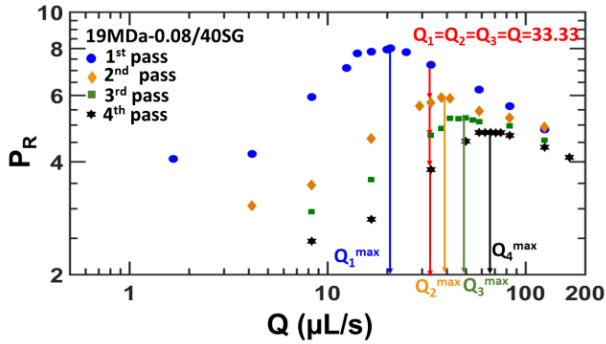

Figure 17: Pressure drop ratio curves of four passes, pre-degraded during every injection at same flow rate $Q_1 = Q_2 = Q_3 = Q = 33.3 \ \mu L/s$.

Figure 17 shows the shifts of the maxima with degradation during successive injections. The history of degradation can be traced by the evolution of maxima from the multiple injection curves. We plot the ratio of the maximum of (N+1)$^{st}$ pass to the maximum of the N$^{th}$ pass *i.e.* Q$_{N+1}$$^{max}$ / Q$_N$$^{max}$, as a function of the normalized flow rate at which it was degraded (collected) during N$^{th}$ pass, Q$_{N+1}$$^{max}$ / Q$_N$$^{max}$ as red stars in Figure 18.

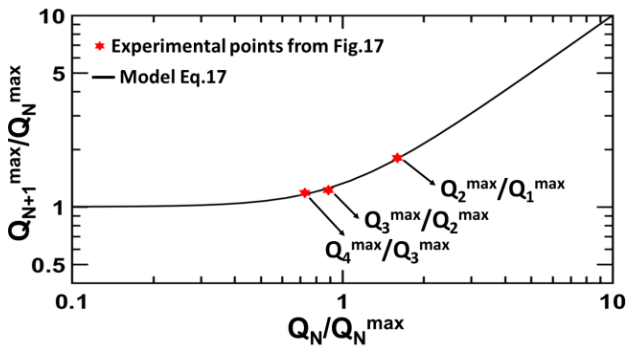

Figure 18: Evolution of the maximum for the data of Fig.17 of a given passage with respect to the maximum of the preceding passage Q$_N$$^{max}$/Q$_{N-1}$$^{max}$, for passages 1, 2, 3, and 4. The model shows a fit to Equation 17 with exponent a=2.5

The points from multiple injections follow the previously obtained trend of for two successive passes shown in Figure 15 (based on Equation 17 with exponent $a = 2.5$). Therefore this relation is capable of predicting the location of the maximum for a pre-degraded polymer solution based on the conditions of degradation ($Q_1, Q_2, Q_3$……. $Q_N$).

Based on this correspondence, we suggest using the entire injection history, in conjunction with the characteristic degradation curve from the first passage, to describe the evolution through multiple passages. The relation between the second and the first passage is given by equation (17). Similarly, the relation between the third and second passages should be given by

$$\frac{Q_3^{max}}{Q_2^{max}} = f\left(\frac{Q_2}{Q_2^{max}}\right) = \left[1 + \left(\frac{Q_2}{Q_2^{max}}\right)^a\right]^{\frac{1}{a}}. \quad (21)$$

Combining equations (17,21) yields

$$\frac{Q_3^{max}}{Q_1^{max}} = f\left(\frac{Q_2}{Q_1^{max}*f\left(\frac{Q_1}{Q_1^{max}}\right)}\right) = \left[1 + \left(\frac{Q_1}{Q_1^{max}}\right)^a + \left(\frac{Q_2}{Q_1^{max}}\right)^a\right]^{\frac{1}{a}}. \quad (22)$$

This can be easily generalized to *N* passages:

$$\frac{Q_N^{max}}{Q_1^{max}} = \left[1 + \sum_{i=1}^{N-1}\left(\frac{Q_i}{Q_1^{max}}\right)^a\right]^{\frac{1}{a}}. \quad (23)$$

In the case of *N* passages at the same flux $Q_i = Q_1$ one finds

$$\frac{Q_N^{max}}{Q_1^{max}} = \left[1 + N\left(\frac{Q_1}{Q_1^{max}}\right)^a\right]^{\frac{1}{a}} \quad (24)$$

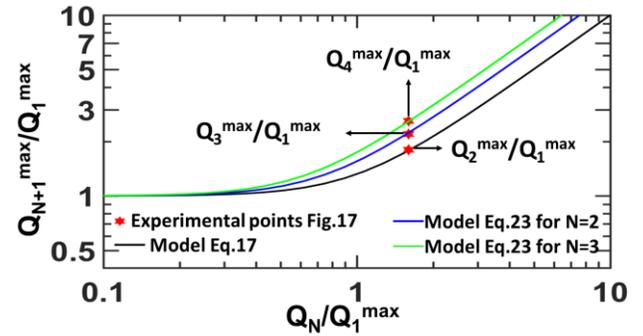

Figure 19: Red stars show the evolution of multiple passages as a function of the 1$^{st}$ pass for the data of Fig. 17. In each case the ordinate is the ratio of Q$^{max}$ for successive injections (1 to 2, 2 to 3, etc.) to Q$^{max}$ of the first injection, and the abscissa is the ratio of Q$^{max}$ of the final injection to Q$^{max}$ of the first injection. The solid lines are calculated using Equation 24 for N=2,3,4 with *a*=2.5.

Eq.23 shows that, with respect to the maximum of the non-degraded solution we observe an upward shift of the maximum flux $Q^{max}$, which corresponds to more scission and lower molecular weight after each pass. Figure 19 shows the iteration of Equation (20) for N = 2,3,4, which agrees well with the measured maxima

## 4. Conclusions

Semi-dilute unentangled polymer solutions just above the overlap concentration were used to study chain scission in a given model contraction. Using a single microfluidic device with a Hencky strain of 3.21, we show that the pressure drop ratio of the polymer solution relative to pure solvent has a maximum as a function of flow rate. The maximum pressure ratio scales with molecular weight and concentration as $P_R^{max} \propto M_w^{1\pm0.2} c^{0.7\pm0.3}$ while the flux at the maximum scales like $Q^{max} \propto M_w^{-1.8\pm0.2} c^{-0.7\pm0.3}$. These scaling are similar to the findings of the other researchers obtained at the onset of degradation [13].

We show that the maximum pressure ratio, which reveals the competition between extensional viscosity and chain scission, occurs for a given Deborah or a given Weissenberg number, although experiments for different Hencky strain are required in order to determine whether such a simple condition applies. We study how the flux at maximum pressure ratio decreases for successive passages through the same contraction, due to different degrees of degradation at each stage. This allows us to quantify the chain scission as a function of the flow rate. We demonstrate a universal pattern of scission that is independent of solvent quality and molecular weight, and depends on the flow geometry (in this case, the Hencky strain). We show that the universal pattern can predict the degradation behavior after successive passes in multiple injection experiments. This robust evolution of scission can be applied in other applications where polymers are injected multiple times, such as lubrication in engines, injection of through ink jets, and during oil recovery; as well as in porous media, which could be represented as successive multiple contractions.

## Acknowledgement

We gratefully acknowledge TOTAL E&P management for permission to publish this work. Thanks to Enric Santanach Carreras and Bertrand Levache for qualitative suggestions and discussions to obtain beautiful and reliable images for data extraction, from the visualization experiments. Thanks to Guenaelle Hauret for help with microfabrication techniques, and to Michele Joly, Cyrille Hourcq and Guillaume Heurteux for support in day-to-day lab activities. PDO is grateful to TOTAL/ESPCI for hospitality and support during his introduction to the project and to the Ives Foundation and Georgetown University for support.

## Bibliography

1. Doi, M. and S.F. Edwards, The theory of polymer dynamics. Vol. 73. 1988: oxford university press.
2. Odell, J.A., et al., Degradation of polymer solutions in extensional flows. Macromolecules, 1990. 23(12): p. 3092-3103.
3. Nguyen, T. and H. Kausch, Mechano-Chemical Degradation of Polymer Solution in Capillary Flow: Laminar and Turbulent Regime. Chimia, 1986. 40(4): p. 129-135.
4. De Gennes, P., Coil-stretch transition of dilute flexible polymers under ultrahigh velocity gradients. The Journal of Chemical Physics, 1974. 60(12): p. 5030-5042.
5. Clasen, C., et al., How dilute are dilute solutions in extensional flows? Journal of Rheology, 2006. 50(6): p. 849-881.
6. Stoltz, C., J.J. de Pablo, and M.D. Graham, Concentration dependence of shear and extensional rheology of polymer solutions: Brownian dynamics simulations. Journal of Rheology, 2006. 50(2): p. 137-167.
7. Haas, R. and W.-M. Kulicke, Characterization of dilute polyacrylamide and polystyrene solutions by means of porous media flow, in The Influence of Polymer Additives on Velocity and Temperature Fields. 1985, Springer. p. 119-129.
8. Keller, A. and J. Odell, The extensibility of macromolecules in solution; a new focus for macromolecular science. Colloid and Polymer Science, 1985. 263(3): p. 181-201.
9. Georgelos, P. and J.M. Torkelson, Apparent thickening behavior of dilute polystyrene solutions in extensional flows. Rheologica acta, 1988. 27(4): p. 369-383.
10. Chow, A., et al., Entanglements in polymer solutions under elongational flow: a combined study of chain stretching, flow velocimetry and elongational viscosity. Macromolecules, 1988. 21(1): p. 250-256.
11. Gupta, R.K., D. Nguyen, and T. Sridhar, Extensional viscosity of dilute polystyrene solutions: Effect of concentration and molecular weight. Physics of Fluids, 2000. 12(6): p. 1296-1318.
12. Müller, A., J. Odell, and S. Carrington, Degradation of semidilute polymer solutions in elongational flows. Polymer, 1992. 33(12): p. 2598-2604.
13. Nghe, P., P. Tabeling, and A. Ajdari, Flow-induced polymer degradation probed by a high throughput microfluidic set-up. Journal of Non-Newtonian Fluid Mechanics, 2010. 165(7-8): p. 313-322.
14. Hunkeler, D., T. Nguyen, and H. Kausch, Polymer solutions under elongational flow: 1. Birefringence characterization of transient and stagnation point elongational flows. Polymer, 1996. 37(19): p. 4257-4269.
15. Hunkeler, D., T. Nguyen, and H. Kausch, Polymer solutions under elongational flow: 2. An evaluation of models of polymer dynamics for transient and

stagnation point flows. Polymer, 1996. 37(19): p. 4271-4281.
16. Odell, J. and A. Keller, Flow-induced chain fracture of isolated linear macromolecules in solution. Journal of Polymer Science Part B: Polymer Physics, 1986. 24(9): p. 1889-1916.
17. Vanapalli, S.A., M.T. Islam, and M.J. Solomon, Scission-induced bounds on maximum polymer drag reduction in turbulent flow. Physics of Fluids, 2005. 17(9): p. 095108.
18. Ryskin, G., Calculation of the effect of polymer additive in a converging flow. Journal of Fluid Mechanics, 1987. 178: p. 423-440.
19. Larson, R. and J. Magda, Coil-stretch transitions in mixed shear and extensional flows of dilute polymer solutions. Macromolecules, 1989. 22(7): p. 3004-3010.
20. Nguyen, T.Q. and H.-H. Kausch, Chain extension and degradation in convergent flow. Polymer, 1992. 33(12): p. 2611-2621.
21. Buchholz, B.A., et al., Flow-induced chain scission as a physical route to narrowly distributed, high molar mass polymers. Polymer, 2004. 45(4): p. 1223-1234.
22. Dupas, A., et al., Mechanical degradation onset of polyethylene oxide used as a hydrosoluble model polymer for enhanced oil recovery. Oil & Gas Science and Technology–Revue d'IFP Energies nouvelles, 2012. 67(6): p. 931-940.
23. Durst, F., R. Haas, and B. Kaczmar, Flows of dilute hydrolyzed polyacrylamide solutions in porous media under various solvent conditions. Journal of Applied Polymer Science, 1981. 26(9): p. 3125-3149.
24. Nguyen, T.Q. and H.H. Kausch, Effects of solvent viscosity on polystyrene degradation in transient elongational flow. Macromolecules, 1990. 23(24): p. 5137-5145.
25. De Gennes, P., Kinetics of collapse for a flexible coil. Journal de Physique Lettres, 1985. 46(14): p. 639-642.
26. Montesi, A., M. Pasquali, and F. MacKintosh, Collapse of a semiflexible polymer in poor solvent. Physical Review E, 2004. 69(2): p. 021916.
27. Durst, F., R. Haas, and W. Interthal, Laminar and turbulent flows of dilute polymer solutions: a physical model, in Progress and Trends in Rheology. 1982, Springer. p. 218-223.
28. Moussa, T., C. Tiu, and T. Sridhar, Effect of solvent on polymer degradation in turbulent flow. Journal of non-newtonian fluid mechanics, 1993. 48(3): p. 261-284.
29. Palangetic, L., et al., Dispersity and spinnability: Why highly polydisperse polymer solutions are desirable for electrospinning. Polymer, 2014. 55(19): p. 4920-4931.
30. Veerabhadrappa, S.K., et al. Polymer screening criteria for EOR application-a rheological characterization approach. in SPE Western North American Region Meeting. 2011. Society of Petroleum Engineers.
31. Rodriguez, S., et al., Flow of polymer solutions through porous media. Journal of non-newtonian fluid mechanics, 1993. 49(1): p. 63-85.
32. Jouenne, S., H. Chakibi, and D. Levitt, Polymer stability after successive mechanical-degradation events. SPE Journal, 2018. 23(01): p. 18-33.
33. Clay, J. and K. Koelling, Molecular degradation of concentrated polystyrene solutions in a fast transient extensional flow. Polymer Engineering & Science, 1997. 37(5): p. 789-800.
34. Dupas, A., Dégradation mécanique de solutions de polymères et ses impacts en récupération assistée d'hydrocarbures. 2012, Université de Bretagne occidentale-Brest.
35. Dobrynin, A.V., Colby, R. H., & Rubinstein, M, scaling theory of poly electrolytes. Macromolecules,, 1995: p. 1859-1871.
36. M.Rubinstein, R.C., Unentangled polymer dynamics, in Polymer Physics. 2003, Oxford: New York. p. 325-330.
37. White, S., A. Gotsis, and D. Baird, Review of the entry flow problem: experimental and numerical. Journal of non-newtonian fluid mechanics, 1987. 24(2): p. 121-160.
38. Boger, D., Viscoelastic flows through contractions. Annual review of fluid mechanics, 1987. 19(1): p. 157-182.
39. Boger, D. and R. Binnington, Experimental removal of the re-entrant corner singularity in tubular entry flows. Journal of Rheology, 1994. 38(2): p. 333-349.
40. Rothstein, J.P. and G.H. McKinley, The axisymmetric contraction–expansion: the role of extensional rheology on vortex growth dynamics and the enhanced pressure drop. Journal of non-newtonian fluid mechanics, 2001. 98(1): p. 33-63.
41. Stelter, M., et al., Investigation of the elongational behavior of polymer solutions by means of an elongational rheometer. Journal of Rheology, 2002. 46(2): p. 507-527.
42. Dupuis, D., et al., Shear thickening and time-dependent phenomena: the case of polyacrylamide solutions. Journal of non-newtonian fluid mechanics, 1994. 54: p. 11-32.
43. Harrington, R. and B. Zimm, Degradation of polymers by controlled hydrodynamic shear1. The Journal of Physical Chemistry, 1965. 69(1): p. 161-175.
44. Edwards, S.F., The theory of polymer dynamics. 1986: Oxford Univ. Press.
45. Rothstein, J.P., and Gareth H. McKinley, The axisymmetric contraction–expansion: the role of extensional rheology on vortex growth dynamics and the enhanced pressure drop. 2001. 98.1;33-63.
46. Rothstein, J.P. and G.H. McKinley, Extensional flow of a polystyrene Boger fluid through a 4: 1: 4 axisymmetric contraction/expansion. Journal of non-newtonian fluid mechanics, 1999. 86(1-2): p. 61-88.
47. Mongruel, A. and M. Cloitre, Axisymmetric orifice flow for measuring the elongational viscosity of semi-rigid

polymer solutions. Journal of non-newtonian fluid mechanics, 2003. 110(1): p. 27-43.
48. Hinch, E., Mechanical models of dilute polymer solutions in strong flows. The Physics of Fluids, 1977. 20(10): p. S22-S30.